\documentclass[letterpaper,useAMS,usenatbib]{mn2e}
\usepackage[totalwidth=500pt, totalheight=680pt]{geometry}
\usepackage{graphicx}

\def\gr{{$\gamma$-ray}}

\def\nup{$\nu_{\rm peak}$}

\newcommand{\lsim}{{\lower.5ex\hbox{$\; \buildrel < \over \sim \;$}}}
\newcommand{\gsim}{{\lower.5ex\hbox{$\; \buildrel > \over \sim \;$}}}
\def\nup{$\nu_{\rm peak}^S$}
 
\newdimen\digitwidth
\setbox0=\hbox{2}
\digitwidth=\wd0
\catcode `#=\active
\def#{\kern\digitwidth}

\newcommand{\fermi}{{\it Fermi}}

\newcommand{\paperone}{{Paper I}}
\newcommand{\papertwo}{{Paper II}}

\title[A simplified view of blazars: the VHE vision]{A simplified view of
  blazars: the very high energy  $\gamma$-ray vision
 }
 \author[P. Padovani and P. Giommi]{P. Padovani$^{1,2}$\thanks{E-mail:
ppadovan@eso.org}, P. Giommi$^{3,4,5}$\\
$^{1}$European Southern Observatory, Karl-Schwarzschild-Str. 2,
D-85748 Garching bei M\"unchen, Germany\\
$^{2}$Associated to INAF - Osservatorio Astronomico di Roma, via Frascati 33,
I-00040
Monteporzio Catone, Italy\\
$^{3}$ASI Science Data Center, via del Politecnico s.n.c., I-00133 Roma Italy \\
$^{4}$ICRANet-Rio, CBPF, Rua Dr. Xavier Sigaud 150, 22290-180 Rio de Janeiro, Brazil\\
$^{5}$Associated to INAF - Osservatorio Astronomico di Brera, via Brera 28, 
I-20121 Milano, Italy\\
}

\begin{document}

\date{Accepted 2014 October 01. Received 2014 September 17; in original form 2014 August 06}

\pagerange{\pageref{firstpage}--\pageref{lastpage}} \pubyear{2014}

\maketitle

\label{firstpage}

\begin{abstract}
  We have recently proposed a simplified scenario for blazars in which these
  sources are classified as flat-spectrum radio quasars or BL Lacs according
  to the prescriptions of unified schemes, and to a varying combination of
  Doppler boosted radiation from the jet, emission from the accretion disk,
  the broad line region, and light from the host galaxy. This scenario has
  been thoroughly tested through detailed Monte Carlo simulations and
  reproduces all the main features of existing radio, X-ray, and \gr\
  surveys. In this paper we consider the case of very high energy emission
  ($E > 100$ GeV) extrapolating from the expectations for the GeV band, which
  are in full accordance with the {\it Fermi}-LAT survey results, and make
  detailed predictions for current and future Cherenkov facilities, including
  the Cherenkov Telescope Array. Our results imply that $\ga 100$ new blazars
  can be detected now at very high energy and up to $z \sim1$, consistently
  with the very recent MAGIC detection of S4 0218+35 at $z=0.944$.
  \end{abstract}


%\begin{document}
%-------------------------------------------------------------------------------

\section{Introduction}\label{intro}

Blazars are a class of Active Galactic Nuclei (AGN) characterised by
distinctive and extreme observational properties, such as large amplitude and
rapid variability, superluminal motion, and strong emission over the entire
electromagnetic spectrum. Blazars also host a jet, pointing almost directly
to the observer, within which relativistic particles moving in a magnetic
field radiate by losing their energy \citep{bla78,UP95}, at variance with
most AGN whose energy production is mostly through accretion of matter onto a
supermassive black hole. 
The two main blazar sub-classes, namely BL Lacertae 
objects (BL Lacs) and flat-spectrum radio quasars (FSRQs), differ mostly in
their optical spectra, with the latter displaying strong, broad emission lines 
and the former instead being characterised by optical spectra showing at most weak 
emission lines, sometimes exhibiting absorption features, and in many cases 
being completely featureless.

Although blazars represent a small fraction of AGN, the interest in this type
of peculiar and rare sources is growing as they are being found in increasing
large numbers in high Galactic latitude surveys performed at microwave and
\gr\ energies \citep{GiommiWMAP09,fermi2lac, PlanckERCSC}.  Blazars represent
also the most abundant extragalactic population at TeV
energies\footnote{http://tevcat.uchicago.edu/}. Very recently
\cite{Pad_2014}, on the basis of a joint positional and energetic diagnostic,
have even suggested a possible association between BL Lacs and
seven neutrino events reported by the IceCube collaboration
\citep{ICECube14}.

In a recent paper \citep[][hereafter Paper I]{paper1} we proposed a new
paradigm, which is based on light dilution, minimal assumptions on the
physical properties of the non-thermal jet emission, and unified
schemes. These posit that BL Lacs and FSRQs are
simply low-excitation (LERGs)/Fanaroff-Riley (FR) I and high-excitation
(HERGs)/FR II radio galaxies with their jets forming a small angle with
respect to the line of sight. We called this new approach the {\it blazar
  simplified view} (BSV). By means of detailed Monte Carlo simulations, Paper
I showed that the BSV scenario is consistent with the complex observational
properties of blazars as we know them from all the surveys carried out so far
in the radio and X-ray bands, solving at the same time a number of
long-standing issues.

In a subsequent paper \citep[][hereafter Paper II]{paper2} we extended the
Monte Carlo simulations to the \gr\ band (100 MeV -- 100 GeV) and found that
our results matched very well the observational properties of blazars in the
\fermi-LAT 2-yr source catalogue \citep[][hereafter
2LAC]{fermi2fgl,fermi2lac} and the \fermi-LAT data of a sample of radio
selected blazars \citep{GiommiPlanck,RadioPlanck}.

%MENTION \cite{pad12}
 
\cite{Arsioli2014} have recently put together a sample of $\sim 1,000$ high
synchrotron peaked (HSP) blazars, that is objects with the frequency of the
synchrotron peak \nup\ $> 10^{15}$ Hz. This was defined starting from a
primary list of infrared (IR) colour-colour selected sources from the
ALLWISE\footnote{http://wise2.ipac.caltech.edu/docs/release/allwise/} survey,
based on data from the Wide-field Infrared Survey Explorer
\citep[WISE;][]{WISE}, and applying further restrictions on IR-radio and
IR-X-ray flux ratios. This so-called WISE HSP (1WHSP) sample is currently the
largest sample of confirmed and candidate HSP blazars. All these objects are
expected to accelerate particles to the highest observed energies and 
radiate up to the very high energy $\gamma$-ray band.
 
The purpose of this paper is to push into the very high energy (VHE; $E >
100$ GeV) band of the electromagnetic spectrum by extrapolating our
simulations from the lower energy (\fermi) \gr\ band and: a) make detailed
predictions for current and future Imaging Atmospheric Cherenkov Telescopes
(IACTs), including the Cherenkov Telescope Array (CTA); b) compare our 
predictions with the number of observed bright 1WHSP sources.
As in \paperone\ and II we use a $\Lambda$CDM cosmology with $H_0 = 70$ km
s$^{-1}$ Mpc$^{-1}$, $\Omega_m = 0.27$ and $\Omega_\Lambda = 0.73$
\citep{kom11}.

\section{Simulations}\label{ingredients}

Our goal is to estimate the properties of a VHE flux-limited blazar sample,
building upon the simulations presented in \paperone\ and \papertwo, applying
the same prescriptions. In \papertwo\ we were mainly interested in
distributions, trends, and average values. To make our predictions as robust
as possible, in view of the required extrapolations, in this paper we
reproduced the absolute numbers in the \fermi\ 2LAC catalogue as well. This
required some minimal fine tuning and a very small number of changes to our
input parameters, which do not have any impact on our previous results.

Our Monte Carlo simulations construct the radio through $\gamma$-ray spectral
energy distributions (SEDs) of blazars at different redshifts and are based
on various ingredients, which include: the blazar luminosity function and
evolution, a distribution of the Lorentz factor of the electrons and of the
Doppler factor, a synchrotron inverse Compton model, an accretion disk
component, the host galaxy. A series of $\gamma$-ray constraints based
on observed distributions estimated using {\it simultaneous} multi-frequency
data are also included: namely, the distribution of Compton dominance, the dependence of the
$\gamma$-ray spectral index on \nup, and that of the $\gamma$-ray flux on
radio flux density. Sources are classified as BL Lacs, FSRQs, or radio
galaxies based on the optical spectrum, as normally done with real
surveys. Readers are referred to \paperone\ and II for full details.

The SEDs were extrapolated to the VHE band by using our simulated \fermi\
fluxes and spectral indices and assuming a break at $E = E_{\rm break}$ and a
steepening of the photon spectrum by $\Delta \Gamma$. Our default values are
$E_{\rm break} = 100$ GeV and $\Delta \Gamma = 1$. The former is consistent
with the SEDs shown in Fig. 8 in \cite{Sent_2013}, while the latter has been
derived as follows: we calculated $\Gamma_{\rm TeV} - \Gamma_{\rm GeV}$ for
the sources in Tab. 1 and 2 of \cite{Sent_2013} with $z \le 0.1$ to minimise
the effect of the extragalactic background light (EBL) absorption. The
resulting value is $\langle \Delta \Gamma \rangle = 1.1$ which, since EBL
absorption is important at higher energies even at low redshifts, should be
taken as a conservative value. To see how strongly our results depend on this
choice we employed also a harder spectrum, that is one with $E_{\rm break} =
200$ GeV and $\Delta \Gamma = 0.5$. VHE spectra were attenuated using recent
estimates of the EBL absorption as a function of redshift \citep{Dom_2011}.

In the rest of the paper we refer to the two different extrapolations as ``soft'' and ``hard'' 
respectively.

\section{Predictions for the VHE band}\label{results}
\begin{figure}
\includegraphics[height=6.1cm]{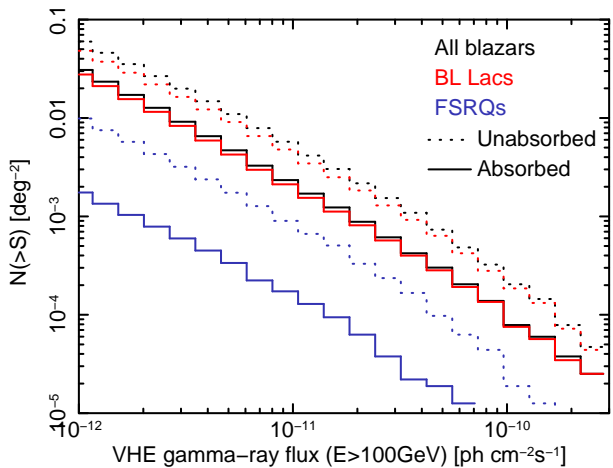}
%\vspace{0.8cm}
\caption{The predicted integral number counts at $E \ge 100$ GeV as a
  function of photon flux with and without EBL absorption (dashed and solid lines respectively) for all blazars
  (black lines), BL Lacs (red lines), and FSRQs
  (blue lines) ($E_{\rm break} = 100$ GeV and $\Delta \Gamma = 1$).}
\label{fig:class_abs_unabs}
\end{figure}

We present the integral number counts as a function of photon flux ($E \ge
100$ GeV) for our default VHE extrapolation ($E_{\rm break} = 100$ GeV and
$\Delta \Gamma = 1$) in Fig. \ref{fig:class_abs_unabs}.

The number counts for the whole blazar sample with and without EBL absorption
are shown as black solid and dotted curves respectively.
We note that the ratio between the two curves is approximately constant and
$\approx 2$, which ultimately stems from our intrinsic spectral slopes and
redshift distributions.

\begin{figure}
\includegraphics[height=8.2cm]{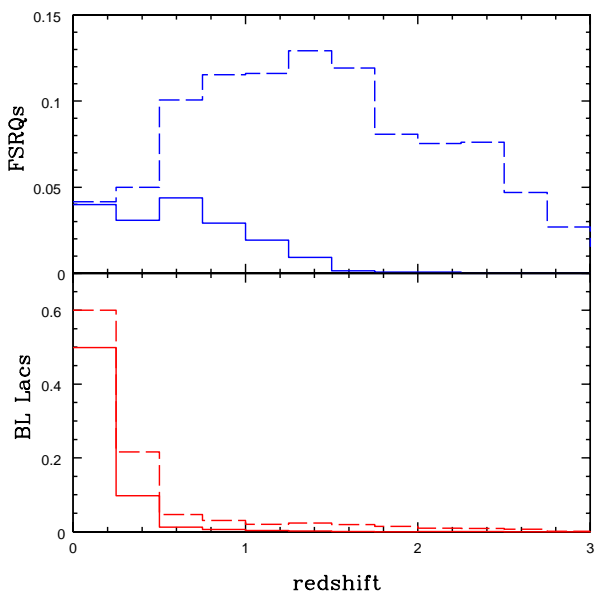}
%\vspace{0.8cm}
\caption{The predicted normalised redshift distributions for FSRQs
  (top panel) and BL Lacs (lower panel) before (dashed lines) and
  after (solid lines) applying the EBL absorption correction ($F (>
  100~{\rm GeV}) \ge 2.5 \times 10^{-12}$ photon cm$^{-2}$ s$^{-1}$,
  $E_{\rm break} = 100$ GeV and $\Delta \Gamma = 1$).}
\label{fig:Nz}
\end{figure}

Fig. \ref{fig:class_abs_unabs} also displays the number counts with and without EBL absorption
for 
BL Lacs (red lines), and FSRQs (blue lines). For  the case of no absorption 
(i.e. assuming a fully transparent Universe) BL Lacs are
$\approx 5$ times more abundant than FSRQs, which is a strong selection
effect to their very different VHE SEDs. 
Since EBL absorption affects FSRQs much more than BL Lacs, due to the larger redshifts of the
former, an even higher ratio ($\approx 15$) between the two classes is expected when absorption is taken into consideration.  
Fig. \ref{fig:Nz} gives a vivid impression of this effect by
showing the redshift distribution of BL Lacs and FSRQs before and after
taking into account EBL absorption (for the case $F (> 100~{\rm GeV}) \ge 2.5
\times 10^{-12}$ photon cm$^{-2}$ s$^{-1}$).  We note that even after the EBL
correction some FSRQs are still detectable up to $z \ga 1.5$ (see below).

\subsection{Current IACTs}

We start by estimating the number of blazars, which are within reach of
present IACTs. This depends on the current typical sensitivity, which is $F
(> 100~{\rm GeV})$\footnote{This value is based on two arguments: 1. the VHE
  detection of 1ES 1312$-$423, which according to \cite{HESS_2013} is ``one
  of the faintest sources ever detected in the very high energy ($E > 100$
  GeV) extragalactic sky'' and, given the fit in the paper, has $F (>
  100~{\rm GeV}) \sim 7 \times 10^{-12}$ photon cm$^{-2}$ s$^{-1}$; 2. Tab. 1
  of \cite{Sent_2013}, where a single source has an integral flux $\sim 1\%$
  of the Crab ($F (> 100~{\rm GeV}) \sim 5 \times 10^{-12}$ photon cm$^{-2}$
  s$^{-1}$), while five others are at a level $\sim 2\%$ of the Crab ($F (>
  100~{\rm GeV}) \sim 10^{-11}$ photon cm$^{-2}$ s$^{-1}$).}$\approx 7 \times
10^{-12}$ photon cm$^{-2}$ s$^{-1}$ ($\sim$ 14 mC.U.)\footnote{1 Crab Unit
  ($E > 100$ GeV) = $5 \times 10^{-10}$ photon cm$^{-2}$ s$^{-1}$ assuming a 
  spectrum $\propto E^{-2.6}$.}. We find
$\sim 330$ blazars all-sky without any EBL absorption, which reduce to $\sim
140$ ($\sim 230$ in the case of the harder VHE spectrum) once EBL absorption is
taken into account. 
Of these, $\sim 91\%$ are BL Lacs 
(7\% FSRQs and 2\% radio galaxies: see below), 
a fraction, which is
similar to the observed one in TeVCat ($94\%$), the
online catalogue for VHE astronomy. Even our predicted fraction of HSP BL
Lacs, $\sim 80\%$, is close to the observed one (86\%). Very recently MAGIC
has detected VHE photons from S4 0218+35, an FSRQ at $z=0.944$, 
during a flaring event \citep{Mirz_2014}. This implies a fraction of FSRQs at $z \ge 0.944$
$=25^{+37}_{-21}\%$, which is also not too far from our prediction of $\sim
10\%$. The total
number, however, is quite different, with only 54 blazars detected so
far. This is easily explained by the fact that TeVCat, although extremely
useful, is only a list of TeV sources that is subject to large biases, as
there are no all-sky flux-limited TeV catalogues at the moment 
\citep[with the
exception of the Galactic plane:][]{2013arXiv1307.4868C}. 
Sources in fact are often pointed by IACTs
when found in a high state in other bands.  
We then predict that $\ga 100$ blazars are waiting to be detected at E~$>$~100 GeV by current IACTs.
For many of these discovery should be relatively easy, as detailed in the next section.

To take into account the fact that the energy threshold of IACTs depends on zenith angle 
we have estimated the fraction of the sky covered by HESS and Veritas/MAGIC 
\cite[Fig. 1 of][where the blind spots corresponds to zenith angles $> 50^{\circ}$]{Dubus_2013}, 
which turned out to be $\sim 90\%$. Our values are therefore overestimated by $\sim 10\%$,
which is well within the range bracketed by our soft and hard extrapolations.

The latest generation of IACTs, (i.e. MAGIC-II and HESS-II) has a lowered energy threshold 
reaching a few tens of GeV, 
with limiting integrated sensitivity of $\sim 2\times10^{-11} $ photon cm$^{-2}$ s$^{-1}$ 
 at E $\ga 50$ GeV \citep{MAGIC-II}. Maximum sensitivity, in terms of 
C.U., is however reached at higher energies: at E~$\ga 300$ GeV the integrated limiting flux is 
$\sim9\times10^{-13} $ photon cm$^{-2}$ s$^{-1}$ \citep{MAGIC-II}.
The number of blazar detections expected by our simulations above these limits 
accounting for EBL absorption
is $\sim 180$ (87\% BL Lacs, 11\% FSRQs) assuming  
E~$\ga 50$ GeV, and $\sim 80$, (96\% BL Lacs, 2\% FSRQs) for the case  
E~$\ga 300$ GeV (these numbers increase by $\sim 20$ and $\sim 210\%$ respectively 
in the case of the harder VHE spectrum). Note that IACTs sensitivities 
are estimated for a Crab-like spectrum, which is not always representative of 
blazars. This, combined with the well known large variability
of this type of AGN, especially just below 100 GeV where the 
steep tail of the GeV emission from FSRQs may play a crucial role, 
makes the above estimates only indicative of average values. IACTs 
observations simultaneous with optical or X-ray flares may lead to a 
larger number of detections.

\subsection{The 1WHSP catalogue}

\cite{Arsioli2014} have recently assembled the 1WHSP catalogue, the largest
sample of confirmed and candidate HSP blazars. Although technically not a
$\gamma$-ray catalogue, it represents at present the best way to compensate
for the lack of large ($|b_{\rm II}| > 20^{\circ}$) sky coverage in the VHE
band for blazars. 1WHSP sources have in fact been shown to be strong
$\gamma$-ray sources: $\sim 1/3$ of them are confirmed GeV emitters and 35
have already been detected in the TeV band. \cite{Arsioli2014} have defined a
``figure of merit'' (FoM) on the potential detectability of their sources in
the VHE band, defined as the ratio between the synchrotron ($\nu f_{\nu}$) peak flux of a
source and that of the faintest blazar in the 1WHSP sample already detected in
the VHE band. Most of the sources with FoM $\ga 1.0$ at low and intermediate
redshifts ($\la 0.7$) should be detectable by the current IACTs, while the majority of
the 1WHSP sample should be within reach of the upcoming CTA. Assuming that the
synchrotron peak flux scales as the VHE flux and for a typical sensitivity
reachable by current IACTs $\sim 14$ mC.U.  we predict $\sim 160$ sources
above such flux and with \nup\ $> 10^{15}$ Hz in the 1WHSP catalogue, which
reduce to $\sim 70$ once the effect of the EBL is taken into account. In the
case of a harder spectrum these numbers almost double. These values need to
be compared with the number of 1WHSP sources with FoM $\ge 1.0$, which is
$\sim 100$, i.e. larger than expected by our default (soft) VHE
extrapolation. The requirements that all sources have an X-ray counterpart
and good WISE photometry in at least three bands, furthermore, imply that the
1WHSP catalogue is somewhat incomplete, which might suggest that the intrinsic
VHE spectrum of blazars is closer to our harder extrapolation than to the
softer one.

\subsection{The CTA case}

We have also made detailed predictions for two distinct hypothetical CTA
surveys: $F (> 100~{\rm GeV}) \ge 2.5 \times 10^{-12}$ photon cm$^{-2}$
s$^{-1}$ (5 mCrab Units [mC.U.]) and a coverage of 10,000 square degrees
(Survey 1: larger area, shallower limit) and $F (> 100~{\rm GeV}) \ge 1.25
\times 10^{-12}$ photon cm$^{-2}$ s$^{-1}$ (2.5 mC.U.) and a coverage of
2,500 square degrees (Survey 2: smaller area, deeper limit). Our results are
presented in Tab. \ref{tab:CTA1_1} -- \ref{tab:CTA2_2} for the unabsorbed and
absorbed cases and for the two different VHE extrapolations.  Based on our
simulations we can draw the following conclusions:

\begin{itemize}

\item the total number of blazars expected in Survey 1 is $\sim 250$ but can
  be even higher ($> 400$) in the case of a harder spectrum. However, the EBL
  reduces the sample to $\sim 110$ objects ($\sim 50\%$ more for the harder
  VHE extrapolation), with $\sim 91\%$ of them being BL Lacs. In Survey 2
  (the deeper but smaller one) we expect slightly less than half the number
  of sources;

\item the mean redshift of the sources gets smaller when EBL absorption is
  applied, which is an obvious consequence of the fact that higher redshifts
  are more affected. What is less obvious is that the mean redshift of FSRQs
  is still $\sim 0.6 - 0.7$, which means that some of these targets can be detected
  up to $z \sim 1.5$, as shown in Fig. \ref{fig:Nz} for the Survey 1
  case. The very recent VHE detection of S4 0218+35 at $z=0.944$ by MAGIC is
  in full agreement with this prediction;

\item the fraction of BL Lacs with redshift increases when the effect of the
  EBL is taken into account from $\sim 70 - 80\%$ to $\ga 90\%$. This is due
  to the fact that the sources with no measurable redshift are those with the
  optical spectra swamped by non-thermal emission, which tend to have a more
  powerful jet and therefore are on average at higher intrinsic
  redshift. Therefore, these are more absorbed than the other BL Lacs;

\item as was the case in \paperone\ and \papertwo\ a small fraction ($\sim 2
  - 5 \%$) of the simulated blazars are classified as radio galaxies. These
  are bona-fide blazars misclassified by current classification schemes
  because their non-thermal radiation is not strong enough to dilute the host
  galaxy component even in the Ca H\&K break region of the optical spectrum
  (see \paperone\ and II for more details).  Being typically at low
  redshifts, these sources are not very much affected by the EBL.

\end{itemize}

\begin{table}
\caption{Simulation of a CTA survey: $F (> 100~{\rm GeV}) \ge 2.5 \times 
10^{-12}$ ph cm$^{-2}$ s$^{-1}$, 10,000 sq. deg.,  $E_{\rm break} = 100$ GeV and
 $\Delta \Gamma = 1$. Mean values correspond to the average of 10 runs.}
\begin{tabular}{llclc}
 Source type & Number & $\langle z \rangle$ & Number & $\langle z \rangle$\\
\multicolumn{5}{l}{~~~~~~~~~~~~~~~~~~~~~~~~~~~~~~Unabsorbed ~~~~~~~~~~~~~~~~~ Absorbed}\\
\hline
BL Lacs     &     203.3 (146.8)$^{a}$   &  0.39 &  103.7 (91.2)$^{a}$  & 0.18\\
FSRQs      &      #40.6       &  1.45  & ##7.1  & 0.62\\
Radio galaxies    &    ##3.6   &  0.05  & ##3.2 & 0.04\\
\hline
Total     &    247.5      &  0.61 & 114.0 & 0.21\\
\hline
\multicolumn{5}{l}{\footnotesize $^{a}$BL Lacs with measurable redshift}
\end{tabular}
\label{tab:CTA1_1}
\end{table}

\begin{table}
\caption{Simulation of a CTA survey: $F (> 100~{\rm GeV})  \ge 2.5 \times 
10^{-12}$ ph cm$^{-2}$ s$^{-1}$, 10,000 sq. deg.,  $E_{\rm break} = 200$ GeV and
 $\Delta \Gamma = 0.5$. Mean values correspond to the average of 10 runs.}
\begin{tabular}{llclc}
 Source type & Number & $\langle z \rangle$ & Number & $\langle z \rangle$\\
\multicolumn{5}{l}{~~~~~~~~~~~~~~~~~~~~~~~~~~~~~~Unabsorbed ~~~~~~~~~~~~~~~~~ Absorbed}\\
\hline
BL Lacs     &     356.7 (277.9)$^{a}$   &  0.41 & 163.5 (147.5)$^{a}$   &  0.19 \\
FSRQs      &      #67.6      &  1.56  & ##8.8 & 0.63\\
Radio galaxies    &    ##8.0  &  0.07  & ##6.2 & 0.06\\
\hline
Total     &    432.3     &  0.62 & 178.5 & 0.21\\
\hline
\multicolumn{5}{l}{\footnotesize $^{a}$BL Lacs with measurable redshift}
\end{tabular}
\label{tab:CTA1_2}
\end{table}
 
\begin{table}
\caption{Simulation of a CTA survey: $F (> 100~{\rm GeV})  \ge 1.25 \times 
10^{-12}$ ph cm$^{-2}$ s$^{-1}$, 2,500 sq. deg., $E_{\rm break} = 100$ GeV and
 $\Delta \Gamma = 1$. Mean values correspond to the average of 10 runs.}
 \begin{tabular}{llclc}
 Source type & Number & $\langle z \rangle$ & Number & $\langle z \rangle$\\
\multicolumn{5}{l}{~~~~~~~~~~~~~~~~~~~~~~~~~~~~~~Unabsorbed ~~~~~~~~~~~~~~~~~ Absorbed}\\
\hline
BL Lacs     &     98.4 (77.6)$^{a}$   &  0.40 & 54.9 (49.6)$^{a}$   & 0.21\\
FSRQs      &      20.1        &  1.52 & #3.4 & 0.71 \\
Radio galaxies    &    #2.6    &  0.07 & #2.3 & 0.06 \\
\hline
Total     &    121.1      &  0.61 & 66.7 & 0.24\\
\hline
\multicolumn{5}{l}{\footnotesize $^{a}$BL Lacs with measurable redshift}
\end{tabular}
\label{tab:CTA2_1}
\end{table}

\begin{table}
\caption{Simulation of a CTA survey: $F (> 100~{\rm GeV})  \ge 1.25 \times 
10^{-12}$ ph cm$^{-2}$ s$^{-1}$, 2,500 sq. deg., $E_{\rm break} = 200$ GeV and
 $\Delta \Gamma = 0.5$. Mean values correspond to the average of 10 runs.}
 \begin{tabular}{llclc}
 Source type & Number & $\langle z \rangle$ & Number & $\langle z \rangle$\\
\multicolumn{5}{l}{~~~~~~~~~~~~~~~~~~~~~~~~~~~~~~Unabsorbed ~~~~~~~~~~~~~~~~~ Absorbed}\\
\hline
BL Lacs     &     162.7 (135.5)$^{a}$   &  0.40 & 82.5 (76.1)$^{a}$   &  0.23\\
FSRQs      &      #31.4      &  1.61  & #4.4  & 0.69\\
Radio galaxies    &    ##6.0   &  0.10 & #4.4 & 0.08 \\
\hline
Total     &    200.1      &  0.60 & 91.3 & 0.24\\
\hline
\multicolumn{5}{l}{\footnotesize $^{a}$BL Lacs with measurable redshift}
\end{tabular}
\label{tab:CTA2_2}
\end{table}

The situation at $E > 1$ TeV is clearly very different. For example, a survey
with $F (> 1~{\rm TeV})$\footnote{This is the approximate unabsorbed limit for Survey 1.}
$\ge 10^{-14}$ photon cm$^{-2}$ s$^{-1}$ and a coverage of 10,000 square
degrees should detect only $\sim 70$ sources taking into account EBL
absorption, $\sim 95\%$ of them being BL Lacs at $z \la 0.3$ and $\sim 1$ source 
being an FSRQ at $z \sim 0.1$. For the harder VHE extrapolation the
total number increases by $\sim 50\%$.

We note that \cite{Dubus_2013} have also predicted the number of blazars CTA
will detect. Comparing our numbers with those in their Fig. 6 we find that,
for example, 
down to $E\times F(> 100~{\rm GeV}) \sim 3 \times 10^{-13} $ erg cm$^{-2}$ s$^{-1}$,
that is, $F (> 100~{\rm GeV}) \sim 1.9 \times 10^{-12}$ photon
cm$^{-2}$ s$^{-1}$, we get $\sim 2 - 3$ (unabsorbed) and $\sim 3 - 4$
(absorbed) times more blazars, depending on the VHE extrapolation. 

This discrepancy is due to the fact that the Dubus et al.'s model is based on 
EGRET data and it is now known to underpredict the faint Fermi source counts 
by a factor of two (Y. Inoue, private communication).

Finally, we stress that our simulated blazar sample does not violate the
$\gamma$-ray background constraint, as measured by \fermi\
\citep{Abdo_2010}. We will address this point in detail in a future paper.

\section{Conclusions}

We have extended our Monte Carlo simulations of the radio through \gr\
emission of blazars within the framework of our proposed blazar simplified
view scenario to the VHE band. The blazar SEDs have been extrapolated under
some simple assumptions and taking into account EBL absorption. Our
predictions are consistent with the number of 1WHSP sources within reach of
current IACTs.  We have also made detailed predictions for two distinct CTA
surveys.

Our main results can be summarised as follows:

\begin{enumerate} 
\renewcommand{\theenumi}{(\arabic{enumi})} 
 
\item current IACTs should be able to detect $\ga 100$ more blazars than
  those already listed in TeVCat. Many of these should be in the 1WHSP
  catalogue;

\item the total number of blazars expected in a CTA survey reaching 5 mC.U.
  at photon fluxes $\ge 100$ GeV, covering 10,000 square degrees, and taking
  into account EBL absorption, is $\sim 110 - 180$, depending on the VHE
  extrapolation. For a survey a factor of two deeper and a factor of four
  smaller we expect about half the number of sources;

\item FSRQs are predicted to make up only a small fraction ($\sim 5\%$) of
  the sources in the CTA surveys we have simulated. However, since their mean
  redshift is $\approx 0.7$, some of them should be detectable up to $z \sim
  1.5$, in agreement with the very recent MAGIC detection of S4 0218+35 at
  $z=0.944$;

\item the redshift of BL Lacs selected at VHE is expected to be easier to
  determine than for BL Lacs in other bands.

\end{enumerate}

\section*{Acknowledgments}

We thank Luigi Costamante, Antonio Stamerra, and an anonymous referee for 
useful input and comments
and Bruno Arsioli for his help in the early phases of this work and for
useful discussions.  PP thanks the ASI Science Data Center (ASDC) for the
hospitality and partial financial support for his visits.

\label{lastpage}
\end{document}